\documentclass[namedreferences]{SolarPhysics}
\usepackage{solaheader}        
\newcommand{\solaaccepteddate}{30 April 2010}
\usepackage[optionalrh]{spr-sola-addons} 
\usepackage{graphicx}        
\usepackage{color}           
\usepackage{url}             

\newcommand{\aap}{    {\it Astron. Astrophys.}}

\newcommand{\aapr}{   {\it Astron. Astrophys. Rev.}}

\newcommand{\apj}{    {\it Astrophys. J.}}
\newcommand{\apjl}{   {\it Astrophys. J. Lett.}}

\newcommand{\apo}{    {\it Appl. Opt.}}

\newcommand{\grl}{    {\it Geophys. Res. Lett.}}

\newcommand{\nat}{    {\it Nature}}

\newcommand{\solphys}{{\it Solar Phys.}}

\newcommand{\caii}     {Ca {\sc ii}~}
\begin{document}
\begin{article}
\begin{opening}
\title{The Mount Wilson \caii K plage index time series\\ {\it Solar Physics}}
\author{L.~\surname{Bertello}$^{1}$\sep
        R.K.~\surname{Ulrich}$^{2}$\sep
        J.E.~\surname{Boyden}$^{2}$
       }
\runningauthor{Bertello, et al.}
\runningtitle{The Mount Wilson \caii K index}
\institute{
            $^{1}$ National Solar Observatory, 950 North Cherry Avenue, Tucson, AZ, USA
            $^{2}$ Department of Physics and Astronomy, University of California, Los Angeles 90095
   email: \url{bertello@noao.edu} email: \url{ulrich@astro.ucla.edu}
   email: \url{boyden@astro.ucla.edu}\\ 
             }
\begin{abstract}
It is well established that both total and spectral solar irradiance are modulated
by variable magnetic activity on the solar surface. However, there is still
disagreement about the contribution of individual solar features for changes
in the solar output, in particular over decadal time scales.
Ionized \caii K line spectroheliograms are one of 
the major resources for these long-term trend studies, mainly because such measurements have
been available now for more than 100 years. In this paper we introduce a new \caii K 
plage and active network index time series
derived from the digitization of almost 40,000 photographic solar images that were
obtained at the 60-foot solar tower, between 1915 and 1985, 
as a part of the monitoring program of the Mount Wilson Observatory. 
We describe here the procedure we applied to calibrate the images and the
properties of our new defined index, which is strongly correlated
to the  average fractional area of the visible solar disk occupied by
plages and active network. We show that the long-term
variation of this index is in an excellent agreement with the 11-year solar
cycle trend determined from the annual international sunspot numbers series. 
Our time series agrees also very well with similar indicators derived
from a different reduction of the same
data base and other \caii K spectroheliograms long-term synoptic programs, such as those 
at Kodaikanal Observatory (India), and at the National Solar Observatory 
at Sacramento Peak (USA). Finally, we show
that using appropriate proxies it is possible to extend this time series up to date,
making this data set one of the longest \caii K index series currently available.
\end{abstract}
\end{opening}
\section{Introduction}

The importance of \caii K spectroheliogram time series for the study
of solar magnetism and irradiance variability is well established. 
Observations of the solar surface reveal magnetic fields
with complex hierarchical structures,
evolving on a wide range of different spatial and temporal
scales. The most prominent aspect of this variability is the solar cycle
of activity, with a period of approximately 11 years for the sunspot numbers and
a period of about 22 years for the magnetic polarity.
Thirty years of satellite measurements of the Sun's energy output have revealed that
also the solar irradiance changes over the full range of time scales from minutes
to decades ({\it e.g.} \opencite{2004A&ARv..12..273F}), 
and this variability is wavelength
dependent \cite{2001GeoRL..28.4119L,1998A&A...329..747S}. 
Empirical models have shown
that the variability in solar irradiance is indeed modulated by the area variations
of the solar surface magnetic features, to a high degree of correlation
\cite{2003AGUFMSH31C..01S,2004ApJ...611L..57F,1988ApJ...328..347F}.
Two of these features, plages and chromospheric
magnetic network, 
account for a significant portion of the Sun's total magnetic flux, UV and EUV variability,
that play a critical role in determining the conditions in the heliosphere which
directly influence the Earth's magnetosphere (see the references in the paper by 
\inlinecite{2009SoPh..255..229F} for a discussion on this subject). 

Observations near the core of the ionized calcium K line (393.37nm) are
one of the most effective tools to investigate the morphology and evolution
of both plages and chromospheric magnetic network. These measurements
have been available since the early years of the twentieth century and, 
because of their correlation to the solar irradiance, they have been widely used 
as proxies to reconstruct the history of the solar magnetism and solar irradiance
over the last 100 years and beyond \cite{2003EOSTr..84..205F,2004ApJ...611L..57F}.
The Mount Wilson archive of ionized \caii K line spectroheliograms provides a
fundamental resource for these studies.
The intensity calibration of these images, however,  is a very difficult task because
the disk brightness and level of background vary from one image
to another due to contributions from the quiet Sun limb darkening curve
of the Ca K-line, the geometrical distortion introduced by the guider errors that
alter the image shape, and the vignetting function produced by the misalignment
of the optical axis with respect to the center of the grating.
For those observations that include the step wedge
densitometer strips it is possible to derive a characteristic curve (also
known as Hurter--Driffield curve, or H\&D curve)
to interpret the transparency of the photographic plate material in terms
of an exposure quantity \cite{1968ApOpt...7.1513D}. Unfortunately, the step wedge
exposures only began 9 October 1961 making this approach unsuitable for the
calibration of the entire database. However, digital filters are convenient tools for
extracting the properties of the spatial intensity distribution of an image.
In this paper we use a median (low-pass) filter to determine the image background 
of each \caii K observation. A flat-fielded image is then obtained
by dividing the image by its background. 
Two of the parameters that describe the pixel distribution of this normalized image 
can be used to define a \caii K plage index and a plage contrast time series.
The very good agreement between these two time series and similar products validates
the analysis described in this paper \cite{2009SoPh..255..229F,2009SoPh..255..239T}.
Finally, we show here that it is possible to extend the 
length of the plage index time series up to date
by using its excellent correlation to the Mount Wilson magnetic plage strength index (MPSI),
making this \caii K index times series one of the longest currently available.

\section{The Mount Wilson Archive}

\begin{figure}
\begin{center}
\includegraphics[width=1.0\textwidth]{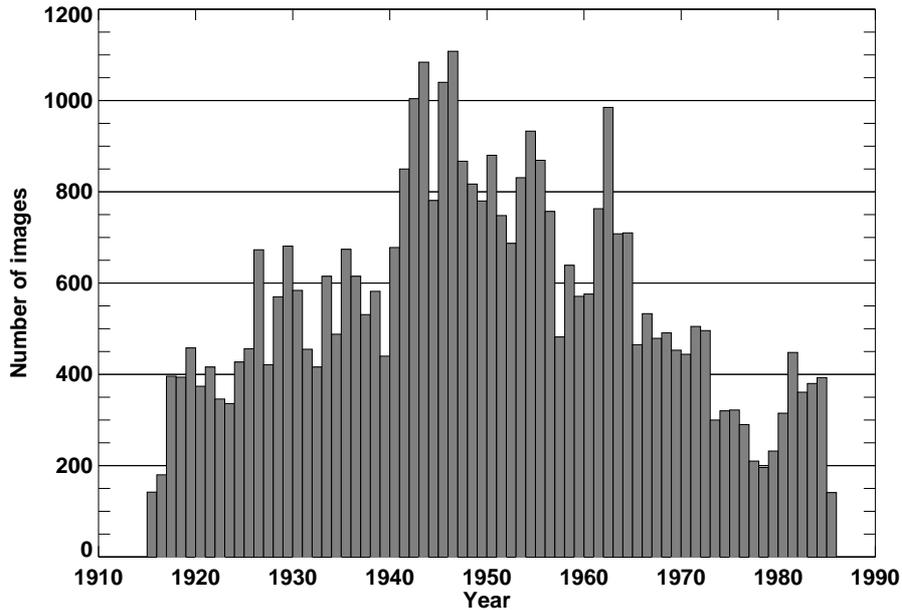}
\caption{Number of solar \caii K spectroheliograms per year investigated by this
study and also available
on line from the project web site. Each image is available as a standard
astronomical format fits file that includes information as to date and time of observation
as well as key data reduction information (see text).}
\label{summary}
\end{center}
\end{figure}

Since the start of the 20th century, the monitoring program of the Mount Wilson Observatory 
has made available to the scientific community a huge number of solar images. 
The glass and acetate negatives are stored and maintained at the Pasadena (California) 
office of the Observatories of the Carnegie Institution of Washington.
This archive contains over 150,000 images of the Sun which were acquired over
a time span in excess of 100 years. The archive includes broad-band images called
White Light Directs, ionized \caii K line spectroheliograms and Hydrogen
Balmer alpha spectroheliograms including both images of the solar disk and images
of prominences above the solar limb. In 2003, the solar physics group at UCLA has
begun a project to digitize essentially all of the \caii K and broad-band direct images
out of the archive, with 12 bits of significant precision and up to 3000 by 3000
spatial pixels. This project has now completed the digitization of the \caii K spectroheliogram
series that began in 1915 and ended in 1985. Almost 40,000 solar images and step 
wedge images (available after 1962) have been extracted and identified with original log-book 
parameters of observation time and scan format. Data from this reduction are accessible from
the project archive on-line at http://www.astro.ucla.edu/$\sim$ulrich/MW\_SPADP.
Figure \ref{summary} shows the distribution of solar \caii K spectroheliograms per year investigated by
this study.

The analysis of this data set will permit a variety of retrospective analyses of
the state of the solar magnetism and provide a temporal baseline of about 100 years
for many solar properties. The chromospheric network is clearly visible on
a substantial fraction of the archive images from the \caii K spectroheliogram
sequence. A previous version of this database was used by \inlinecite{2001GeoRL..28..883F}
to study the quiet network contribution to solar irradiance
variations. These variations are important because they can be used
to investigate the influence of the solar luminosity change on the Earth's
climate \cite{2006Natur.443..161F}. More recently, the digitized images obtained
from this archive were independently calibrated by \inlinecite{2009SoPh..255..239T}
along with \caii K data from Kodaikanal Observatory and from the National Solar
Observatory at Sacramento Peak, for the purpose of comparing the signature
of plages and enhanced network. The comparison shows a good agreement in
the temporal behavior of the annual and monthly mean plage indices calculated from
the three data sets. A more extended comparison, that includes six time series of
annual mean plages indices, shows a similar result \cite{2009SoPh..255..229F}.
The image quality and contents of the \caii K spectroheliogram time series 
obtained by the digitization of the Arcetri, Kodaikanal, and Mt Wilson photographic archives
have been also evaluated by \inlinecite{2009ApJ...698.1000E} to estimate
their value for studies focusing on timescales longer than the solar cycle. 

\section{Data Analysis}
\subsection{Pre-Processing and Calibration of the Images}

The procedures used to calibrate the \caii K images are described
in some details at the project web site www.astro.ucla.edu/$\sim$ulrich/MW\_SPADP.
We only provide here a short summary of the most important steps included in this
calibration process, relevant to the analysis described in this work. 
The images present notably some dust and pit which is
important to reduce. A Laplacian filter was used for this purpose. The size of
the images was then reduced from its original scanned resolution of $\approx 3000 \times 3000$
to $\approx 866 \times 866$ spatial pixels, by averaging the pixel values within
each 4$\times$4 portion of the image. 
Due to the nature of the scans, where the dispersion direction is in the direction of the scan and 
the cross-dispersion direction is parallel to the slit, a strategy was adopted whereby the search in 
the image is for radii associated with the scan/dispersion direction
and cross-dispersion/slit direction independently. 
Because the average distortion of each image
amounts to less than $\approx$0.9\%, with the radius along the scan/dispersion direction
being typically smaller than the radius along the cross-dispersion/slit direction,
an average value of the radius was used. 
Due to small changes in the image-scale size over the years, the annual mean value of this
average radius varies from about 336 pixels to 340 pixels, with not significant
long-term trends over the 70-year period investigated by this study.

\begin{figure}
\begin{center}
\includegraphics[width=1.0\textwidth]{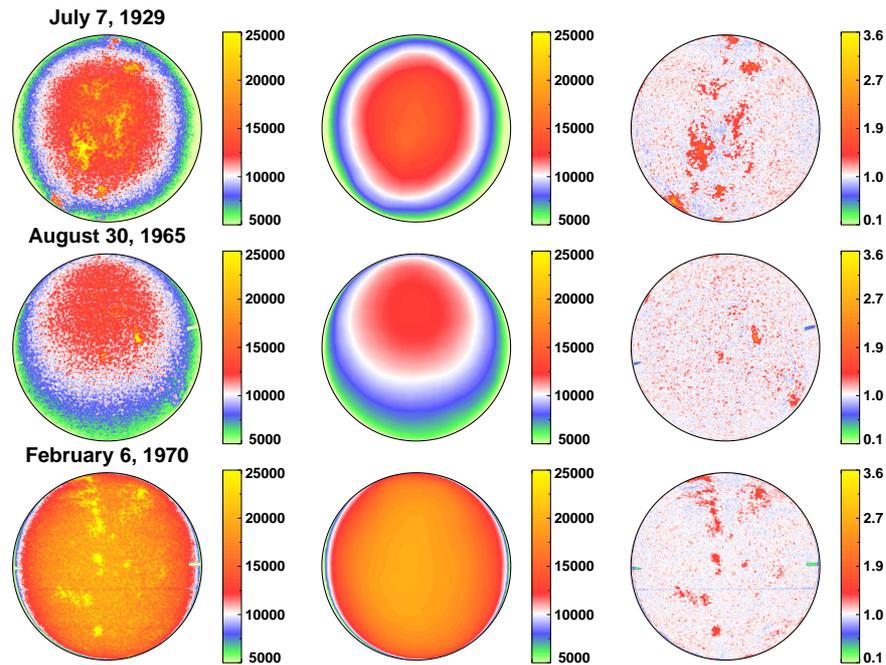}
\caption{Examples of median-filtered images. The left column shows the original
images taken on the days indicated in the figure. The middle column displays
the background of the corresponding image calculated using a running median filter,
essentially a low-pass filter, and the last column is the result of the flat-fielding
obtained by dividing the previous two images together. The chromospheric network is
well defined on the final images. Also, on the last two images, the polar marks indicating
the solar north-south direction are clearly visible. 
These marks are present however only on a portion of the dataset.}
\label{median}
\end{center}
\end{figure}

One of the main problems we encountered during
the calibration process is the presence of a vignetting function. This function
is linked to the relative position between the pupil and the grating,
which depends on the coelostat mirror positions and shifts during each exposure
due to the scanning of the spectrograph across the solar image. As a result
of this effect the intensity and its gradient are highly variable from one image
to another. Stepwedge exposures on the \caii K spectroheliograph sequence are available
only since 9 October 1961. For those images it is possible to use, 
following the ideas presented by \inlinecite{1968ApOpt...7.1513D}, the H\&D
curve calibration approach to obtain well calibrated intensity images. Unfortunately,
the bulk of the sequence does not offer this possibility and a different approach
is required. In our study we used a running median filter to determine the large-scale spatial
intensity distribution of the image, i.e. its background. We then divide the image by 
this background
to obtain the final flat-fielded image. Figure \ref{median} illustrates the procedure for
a sample of three different observations: The first column shows the original images after
they have been corrected for dust and pit and reduced in size. The variation in intensity
is clearly visible on all these images. Moreover, the dynamic range is also quite different
from one image to another being, for example,  much lower in the last image. The
second column shows the background determined from the running median filter, and the
last column is the flat-fielded image obtained by dividing the first two images. The parameters
of the median filter have been accurately determined by comparing the result of the filter to
the proper H\&D calibration for images taken after 9 October 1961. This is an important
and necessary step to avoid introducing spurious points through an improper
treatment of the non-linear response curve that could affect the values of the
CaK index defined in the next section. The pixel value of the background in the
flat-fielded image is typically around unity.

\subsection{The CaK Index Series}

A CaK index is defined from each flat-fielded image using a multi-step
procedure based on the distribution of the intensity ratio, as
described in details below.

\begin{figure}
\begin{center}
\includegraphics[width=1.0\textwidth]{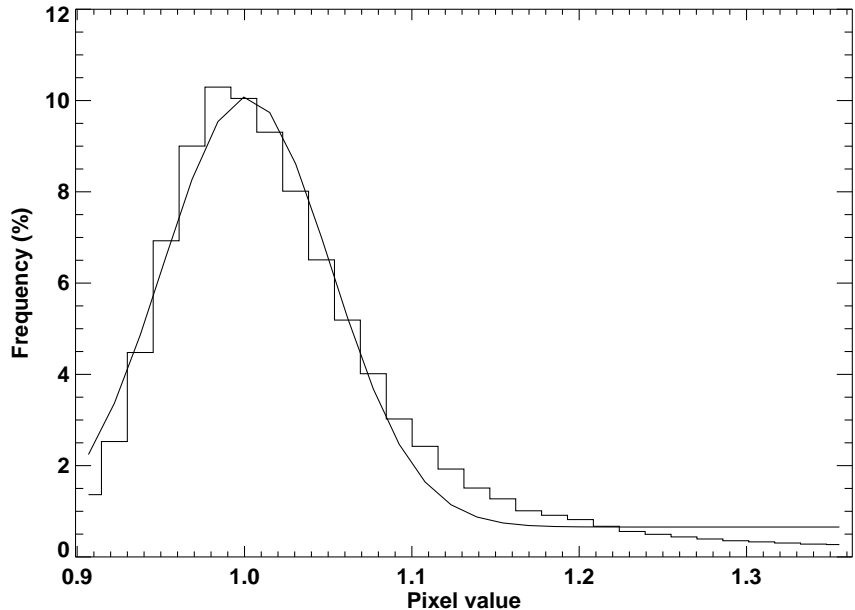}
\caption{The intensity-ratio distribution for an image taken 2 July 1970. The fundamental
parameters of this distribution are determined from a Gaussian fit to the histogram, which 
contains 30 bins between $x_{\rm c} - 2\sigma$ and $x_{\rm c} + 7\sigma$. This is the
exact range shown in the plot, where $x_{\rm c}$ and $\sigma$ are
the center and the width of the distribution, respectively. Also shown is the best
Gaussian fit to the distribution. The constant baseline $B$ (see Equation (1)) defines the
CaK plages and active network index used in this work.}
\label{histo}
\end{center}
\end{figure}

\begin{enumerate}
\item A first histogram is calculated using all pixels located within 0.99 solar radii from
the center of the image and with values less than 3.
The binsize of the histogram is chosen to be 0.01. The width $\sigma$ and the center 
$x_{\rm c}$ ($x_{\rm c} \approx 1$) of this 
distribution are determined from a Gaussian fit to the histogram using an equal number
of bins (35) around the maximum of the distribution. This choice of parameters limits
the analysis to the central part of the histogram, which is well described by a Gaussian
function.
\item The calculated width $\sigma$ is used to define the boundaries of a new histogram,
between $x_{\rm c} - 2\sigma$ and $x_{\rm c} + 7\sigma$. 
To properly account for the contribution of both plages and network to the skewed distribution
shown in Figure \ref{histo}, the long tail on the right side of the histogram needs to be included in the fit.
We find that a value of 7$\sigma$ for the tail accomplishes this goal and provides results that are in
excellent agreement with other studies as shown in this paper.
This range
is then divided into 30 bins, so that the binsize varies in general
from one distribution to another. The histogram is then normalized by dividing the value of
each bin by the total number of pixels in the solar disk image.
This normalization is necessary, to properly
take into account the change in the observed solar disk area produced by the
variable Sun-Earth distance.
\item The following four-parameter Gaussian function is used to model the distribution: 
\begin{equation}
 y(x) = A\exp(-u^2/2) + B,
\end{equation}
where $u = (x - x_{\rm c})/\sigma$, $x$ is the bin value, and $y$ is the fractional number of
pixels in the solar disk with value $x$. 
\end{enumerate}
Figure \ref{histo} shows a typical histogram derived from a single observation, defined between
the boundaries $x_{\rm c} - 2\sigma$ and $x_{\rm c}+ 7\sigma$. The solid curve is the Gaussian
best fit to the distribution given by Equation (1) It is quite
evident from the figure that the distribution is well described by a Gaussian function
only for
pixel values around to the center ($x_{\rm c} \approx 1$) of the histogram, and tends to drift
away from it as the pixel value increases. Since relatively high intensity-ratio values
correspond to brighter than average features in the image, that is plages and enhanced network, 
the analysis of Figure \ref{histo} shows that those features will add contributions to the right
side of the histogram and make the distribution asymmetric.
The measurement of the constant baseline $B$ from Equation (1) provides an average
estimate of this effect.

\subsection{The \caii K Plages and Active Network Index Series}

\begin{figure}
\begin{center}
\includegraphics[width=1.0\textwidth]{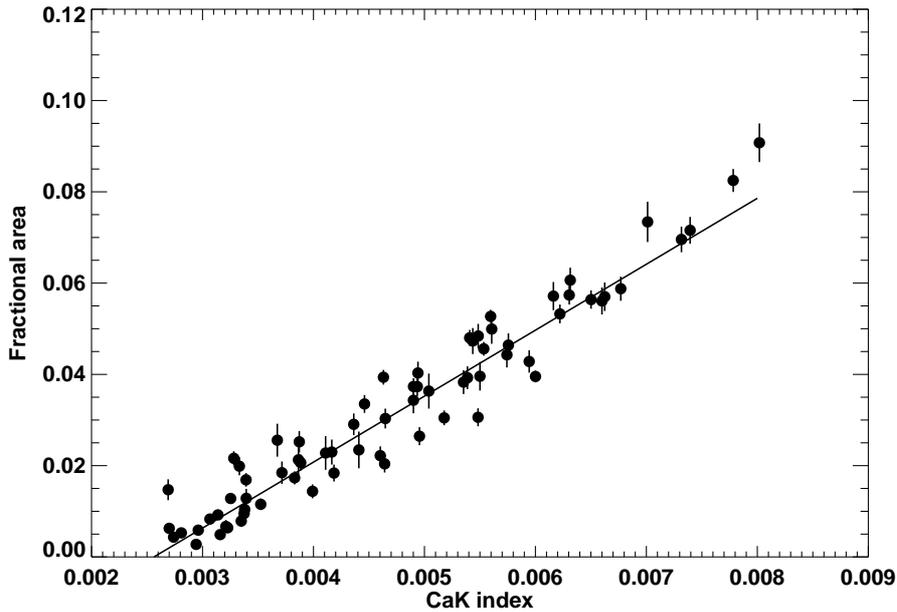}
\caption{Comparison between the annual means of two indices derived from independent
analyses of the Mount Wilson archive of \caii K observations. The CaK index is the value
measured according to the procedure described in this paper, while the fractional areas
covered by plages and active network were measured by 
Tlatov, Pevtsov, and Singh (2009).
The annual mean areas were calculated from the monthly values provided by the 
authors and the error bars are the 1-sigma uncertainties of the mean. For the CaK index,
the error bars are much smaller than the size of the points showed in the plot. Also
shown is the regression line, given by Equation (2)}
\label{areas}
\end{center}
\end{figure}

The value of the constant baseline $B$ in Equation (1) is defined in this work to be an index of 
the \caii K plages and active network, or simply a CaK index. It is important to notice that
despite the normalization introduced in this definition, our index does not directly
measure the fraction of the solar hemisphere occupied by the chromospheric plages and
enhanced network. Instead, it measures the average effect of those features on the intensity-ratio
distribution of the image. 
As discussed in Section 3.1, the lack of information about the characteristic curve
for most of the images in the sequence, in addition to the unknown intensity contribution
from scattered light, makes the task of obtaining a homogeneous and consistent series
of well calibrated images very difficult. Consequently, the calculation of the actual area
of the solar disk covered by plages and/or network must rely on some additional assumptions
about the physical properties of the database.
In their independent analysis of this series, \inlinecite{2009SoPh..255..239T} have
derived a CaK index expressed in units of fraction of the solar hemisphere covered 
by plages and network on the assumption that the center-to-limb intensity
variation in quiet Sun corresponds to a standard curve, independent of overall level
of solar activity. This result can be used to calibrate our time series on the same
scale. 
Figure \ref{areas} shows the
comparison between our annual means and their corresponding values. 
The linear Pearson correlation coefficient of these two indices is 0.97, 
with a Student's t-test showing a significant correlation between the two
variables at a confidence level much higher than 99.9\%,
indicating the 
excellent agreement of the two series. The linear regression is described by the equation
\begin{equation}
\mbox{Fractional area} = -3.70\times10^{-2}(\pm 0.06\times10^{-2}) + 14.45(\pm
0.15)\mbox{CaK index}.
\end{equation}
This equation can be used to convert our CaK index into a fractional area, so that the
comparison with similar time series can be made. 

\begin{figure}
\begin{center}
\includegraphics[width=1.0\textwidth]{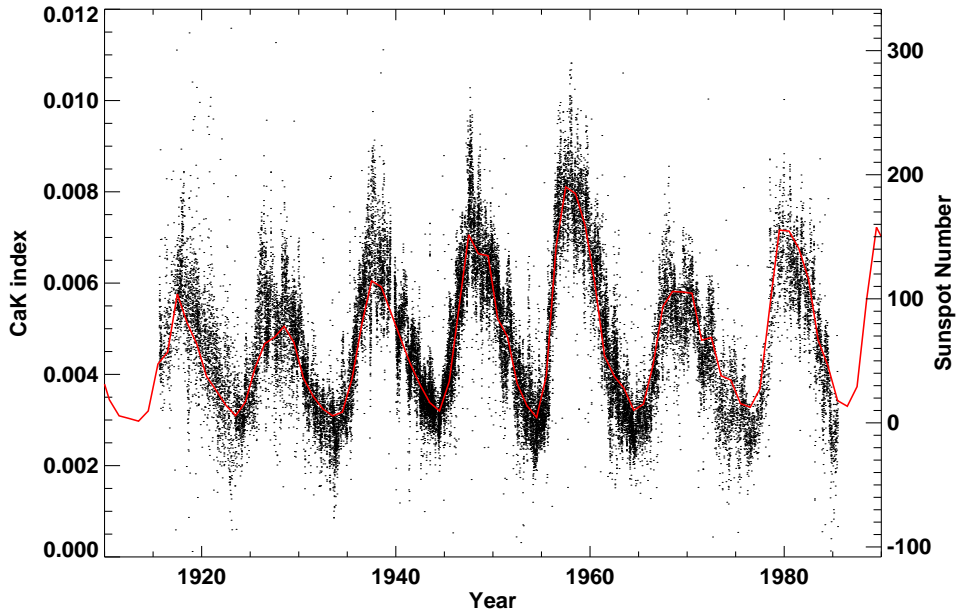}
\caption{The CaK index  time series defined using the
fitting procedure described in the text. This index measures the contribution
from both plages and active network. 
Almost 40,000 images have been reduced to
produce this plot. The annual international sunspot numbers are also shown, in red, as a reference.
}
\label{cak}
\end{center}
\end{figure}

We have applied the procedure described in the previous section to the entire dataset of
observations taken between 1915 and 1985. The value determined from each image 
is plotted in Figure \ref{cak}, together with the temporal annual
variations of the international sunspot numbers. No attempt has been done to eliminate
any specific observations from this analysis. From the visual inspection
of the figure it is quite evident that our CaK index well correlates with
the 11-year solar cycle defined by the sunspot numbers. In addition, comparisons between
our CaK annual mean values  and other available CaK plage indices
shows that our definition is consistent with those based on the area of plages
including the active network \cite{2009SoPh..255..229F,2009SoPh..255..239T}. 

\begin{figure}
\begin{center}
\includegraphics[width=1.0\textwidth]{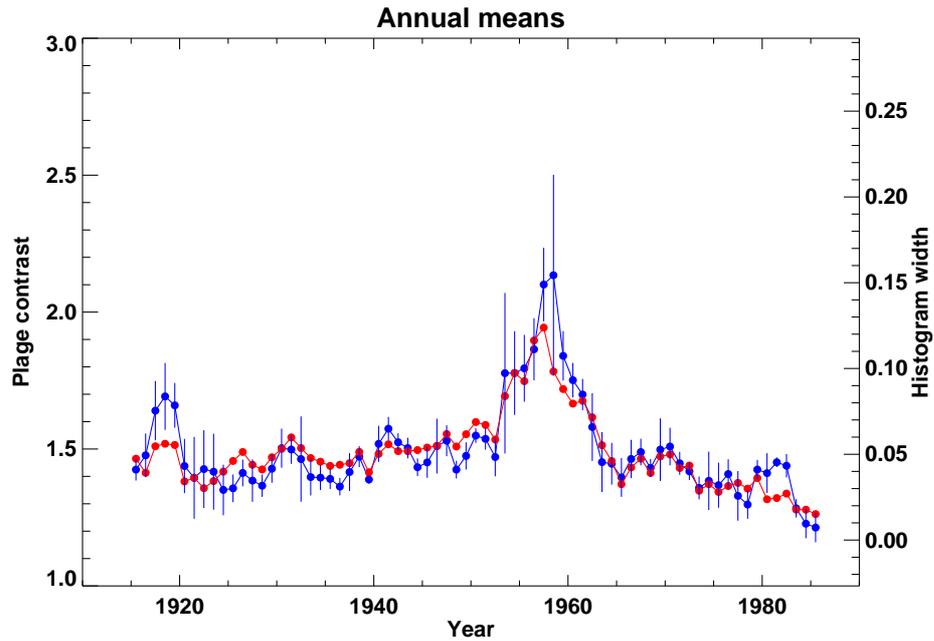}
\caption{Comparison between the plage contrast (blue) measured by 
Tlatov, Pevtsov, and Singh (2009) and
the annual mean width values (red) derived from the histogram analysis 
described in this paper. The error bars are 3 standard deviations of the mean. In our
analysis the error bars are much smaller, smaller than the size of the individual (red)
points, mainly because of the much large sample used in the investigation.}
\label{contrast}
\end{center}
\end{figure}

In addition to the CaK index time series, defined by the parameter $B$, 
the fit to the histogram provides supplementary information about the properties
of the distribution in intensity of each image. In particular, the behavior
of the width $\sigma$ seems to be correlated to the plage contrast. In their
investigation of the Mount Wilson \caii K photographic archive \inlinecite{2009SoPh..255..239T}
defined a plage contrast as the plage brightness per
unit area. Figure \ref{contrast} shows the comparison between $\sigma$ and
their plage contrast. Although the two quantities have
been defined differently and the calibration of the photographic plates was done
using completely separated methods, the two curves show very similar variations over
time. The plage contrast time series seems to be more sensitive to the variations
in the cycle of activity than the result obtained from the histogram analysis. 
The most interesting feature of this figure is the increasing in the plage contrast
during solar cycle 19, from about 1953 to 1961. A similar analysis performed by
\inlinecite{2009SoPh..255..239T} on the archive of the Kodaikanal Observatory (India) for the period
1907-1999 does not show any significant long-term variation of the plage contrast.
It is speculated that the increase in the plage contrast during solar cycle 19 for
the Mount Wilson data is due to the exit slit width being narrower during that 
period \cite{2009SoPh..255..239T}. 
In fact, an important property of the images obtained during
this interval is the present of dark filaments which are known to be prominences
projected onto the solar disk. These features are common for K3 spectroheliograms
but not for K2 spectroheliograms, the bulk of this database. 
The interval for which the filaments are evident coincides closely with this period of 
anomalous plage contrast.

\subsection{The Extended CaK Index Series}

\begin{figure}
\begin{center}
\includegraphics[width=1.0\textwidth]{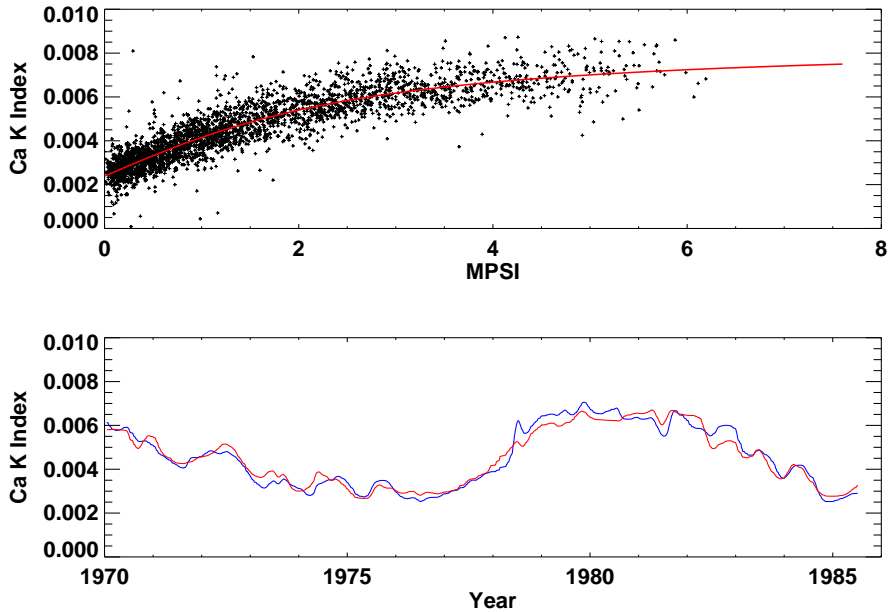}
\caption{Relationship between the daily CaK index values and MPSI for the time
interval 1970-1985. Only days for which both quantities were available have been
included in this analysis. This relationship can be described with a simple
three-parameter model as described in the text.
The solid red line shows the best fit to the data (top plot).
A comparison between the temporal variations of the calculated CaK index (blue curve) and
the rescaled MPSI (red curve), smoothed using a 60-day width running Gaussian, 
are also shown (bottom plot).
}
\label{mpsi}
\end{center}
\end{figure}

Since 1970, the synoptic program of Doppler and magnetic observations at the 150-foot
solar tower at Mount Wilson has provided to the solar community a 
MPSI extracted from the Fe {\sc i}~ 5250 \AA~  magnetograms.
The MPSI is defined as the
sum of the absolute values of the magnetic field strengths for 
all pixels where the absolute value of the magnetic field strength is between 10 and 
100 gauss \cite{1986ApJ...302L..71C}.
This number is then divided by the total number of pixels 
(regardless of magnetic field strength) in the magnetogram. 
The entire record of MPSI daily measurements is available from
www.astro.ucla.edu/$\sim$obs.
As shown in figure
\ref{mpsi}, our daily CaK index is well correlated with the MPSI during the time interval 
from 1970 to 1985. In particular, the top plot shows a quite strong linear correlation
up to a MPSI value of about 2, but then this correlation decreases for larger
MPSI numbers. The overall relationship can be well described with a simple
three-parameter model, given by $a + b\arctan(c\cdot{\rm MPSI})$, where $a, b,$ and $c$
are the three parameters to  be determined using the method of least squares.
Our best fit to the data is shown by the solid red line in top plot of figure \ref{mpsi}.
The lower plot shows the temporal behavior of the
rescaled MPSI (red line) compared to the corresponding CaK quantity. The MPSI
was rescaled using the coefficients from the above model. 
The comparison clearly shows the excellent agreement between the two variables.

\begin{figure}
\begin{center}
\includegraphics[width=1.0\textwidth]{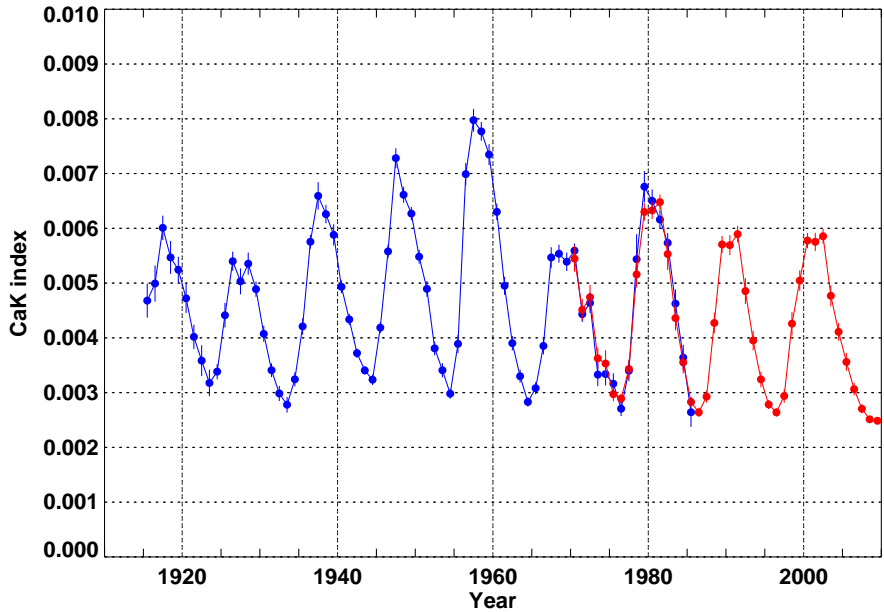}
\caption{The extended CaK index series. In blue are the 
annual mean values obtained from the original spectroheliograms, while
in red are the values derived from the MPSI calibrated into the CaK scale.
The 16 years overlap between the two series, from 1970 to 1985, show the excellent
correlation between the two data sets and validate the use of the MPSI as a proxy
to extend the CaK index data. The error bars are 5 times the standard
deviation of the mean.
}
\label{xcak}
\end{center}
\end{figure}

An important consequence of this excellent correlation is the possibility of using the MPSI
as a proxy for the CaK index, and therefore extending the calculation of
this index up to date. Using the coefficients derived from the model shown in 
the top plot of figure \ref{mpsi} we were able to update the series up to 13 July
2009. The annual mean values for the period 1915-2009 are shown in figure \ref{xcak},
where the blue points are obtained from the 1915-1985 original spectroheliogram images 
and the red points from the 1970-present MPSI measurements. These values are also
listed in table 1. The most significant feature of this figure, and table 1,
is the fact that the value of our
CaK index during periods of minimum solar activity has slightly but systematically 
decreased since 1940. In particular, our analysis seems to suggest that the current 
minimum has produced the lowest CaK index value among those in the almost hundred
years investigated by this study. 

\begin{table}
\caption{The annual averages of the CaK time series and the corresponding
standard deviation (rms) of the mean. The values after 1985 have been
derived from the MPSI time series, used as a proxy (see text).}
\begin{tabular}{lllllllll}
\hline
Year & Index & rms & Year & Index & rms & Year & Index & rms \\
  & $\times10^{-5}$ & $\times10^{-5}$ &   & $\times10^{-5}$ & $\times10^{-5}$ & & $\times10^{-5}$ & $\times10^{-5}$ \\
\hline
1915&
  467.98&
 6.22&
1947&
  728.16&
 3.62&
1979&
  675.98&
 5.69\\
1916&
  499.23&
 6.60&
1948&
  661.32&
 2.94&
1980&
  650.16&
 4.18\\
1917&
  600.91&
 4.44&
1949&
  626.94&
 2.46&
1981&
  615.96&
 3.39\\
1918&
  546.83&
 6.01&
1950&
  548.21&
 2.54&
1982&
  573.47&
 3.55\\
1919&
  524.30&
 4.78&
1951&
  489.41&
 2.90&
1983&
  462.49&
 5.21\\
1920&
  472.13&
 5.33&
1952&
  380.80&
 2.53&
1984&
  364.02&
 4.65\\
1921&
  401.88&
 4.47&
1953&
  340.81&
 2.58&
1985&
  272.24&
 3.08\\
1922&
  358.41&
 5.62&
1954&
  298.11&
 1.84&
1986&
  264.05&
 1.68\\
1923&
  317.68&
 4.89&
1955&
  389.04&
 3.31&
1987&
  292.68&
 2.19\\
1924&
  338.34&
 2.80&
1956&
  698.96&
 4.05&
1988&
  427.24&
 3.76\\
1925&
  441.29&
 4.46&
1957&
  797.44&
 4.15&
1989&
  570.57&
 3.10\\
1926&
  539.94&
 3.50&
1958&
  777.01&
 3.43&
1990&
  569.22&
 3.66\\
1927&
  503.16&
 4.71&
1959&
  734.62&
 3.82&
1991&
  589.40&
 3.00\\
1928&
  535.43&
 4.05&
1960&
  630.05&
 3.03&
1992&
  485.43&
 4.73\\
1929&
  488.86&
 2.79&
1961&
  495.05&
 3.29&
1993&
  395.17&
 3.60\\
1930&
  407.35&
 2.90&
1962&
  390.19&
 2.63&
1994&
  324.17&
 2.89\\
1931&
  340.92&
 2.52&
1963&
  329.97&
 2.33&
1995&
  278.37&
 1.67\\
1932&
  298.51&
 2.76&
1964&
  282.97&
 1.70&
1996&
  263.90&
 1.50\\
1933&
  277.89&
 2.85&
1965&
  308.19&
 2.19&
1997&
  293.92&
 2.52\\
1934&
  324.32&
 2.57&
1966&
  385.17&
 3.07&
1998&
  425.89&
 4.23\\
1935&
  420.96&
 3.00&
1967&
  546.75&
 3.77&
1999&
  504.58&
 3.74\\
1936&
  575.43&
 2.53&
1968&
  553.61&
 3.30&
2000&
  577.69&
 2.74\\
1937&
  659.24&
 5.00&
1969&
  538.95&
 3.38&
2001&
  575.31&
 3.30\\
1938&
  625.86&
 3.38&
1970&
  559.24&
 2.59&
2002&
  585.56&
 2.75\\
1939&
  588.09&
 3.86&
1971&
  443.25&
 2.84&
2003&
  476.82&
 3.85\\
1940&
  493.13&
 2.35&
1972&
  463.73&
 3.72&
2004&
  410.99&
 3.23\\
1941&
  433.68&
 2.06&
1973&
  332.84&
 4.18&
2005&
  356.04&
 3.24\\
1942&
  372.09&
 1.72&
1974&
  333.89&
 4.17&
2006&
  306.31&
 2.29\\
1943&
  340.59&
 1.53&
1975&
  316.23&
 3.81&
2007&
  270.50&
 1.63\\
1944&
  323.59&
 1.95&
1976&
  270.64&
 2.67&
2008&
  251.31&
 1.02\\
1945&
  418.65&
 2.03&
1977&
  339.96&
 3.78&
2009&
  248.60&
 0.69\\
1946&
  557.84&
 2.48&
1978&
  543.58&
 9.13&
 & & \\
\hline
\end{tabular}

\end{table}

\section{Discussion and Conclusions}

In this paper we have described a method to study the time dependence of
solar surface magnetism from the analysis of more than 70 years
of \caii K spectroheliograms obtained at the 60-foot solar tower in Mount Wilson
between 1915 and 1985. These measurements are important because \caii K emission
is closely related to the magnetic field \cite{1998SoPh..177..265J}
and can provide a proxy for long-term
induced changes, solar-cycle and beyond, in the total and spectral solar irradiance
({\it e.g.} \opencite{2000SSRv...94..139F}).
Our method is based on the photometric properties of
each individual solar images, using parameters computed
from a histogram analysis of their intensity distribution. 
An important feature of this approach is the fact that it can be accomplished in a fully
automated mode, without relying on a visual inspection of the images. 
We found that two of the parameters defining this intensity distribution,
the constant baseline and the width, are very
well correlated to the fractional area of the solar disk covered by plages plus
active network and the plage contrast as computed by an independent investigation
of the same database. The constant baseline measures the strength of the
widely distributed regions that are brighter than average in \caii K, and it
is defined in this work as a CaK index. This index
shows a temporal behavior which is in excellent agreement with the cycle of solar
activity described by the international sunspot number. This is clearly illustrated 
in Figure \ref{cak}, where the relative strength of solar cycles 15 to 21 is well
reproduced.

Another remarkable property of this index is its correlation with the MPSI measurements
shown in Figure \ref{mpsi}. The relationship between these two variables is mostly
linear, up to a MPSI value of about 2. Above this value the correlation drops
quite significantly. Those are measurements corresponding to the time interval from 1979
to 1982.4, around the maximum of solar cycle 21, where both the MPSI and the
CaK index show no systematic trends. As illustrated at the bottom of Figure \ref{mpsi}, our
adopted model produces an excellent agreement between these two quantities during both
the period of minimum of activity, around 1975, and the maximum of cycle 21. 
The validity of this model, however, is limited to MPSI values up to about 7. In fact,
above this limit, the asymptotic behavior of the model significantly reduces the correlation
between these two indices of solar activity. Fortunately,
an investigation of the historical record of MPSI daily measurements
shows that this event is extremely rare, and will not affect the conclusions reached
in this paper.

The possibility of extending the CaK index series beyond the time for 
which the \caii K spectroheliograms
are available, using the MPSI measurements as a proxy, is one of the main results of this work.
This unique extended series, that covers almost a century of measurements, is shown in Figure
\ref{xcak} and tabulated in Table 1. Because of the close connection among
\caii K emission, solar surface magnetic field, and Total Solar Irradiance (TSI), 
this CaK time series can provide
a proxy for studies on solar-cycle induced changes in the TSI that may play a role in climate
change \cite{2002AdSpR..29.1933S,2000SSRv...94..337S}. In particular, 
the progressive increase in the strength of the solar cycle of activity between 1928
and 1958 indicated by the sunspot number is closely tracked by the increase in strength
of solar UV radiation as indicated by our CaK index. This probable increase in the solar
output during a period of increasing greenhouse gas abundance needs to be included in
evaluations of the global warming process.

\begin{figure}
\begin{center}
\includegraphics[width=1.0\textwidth]{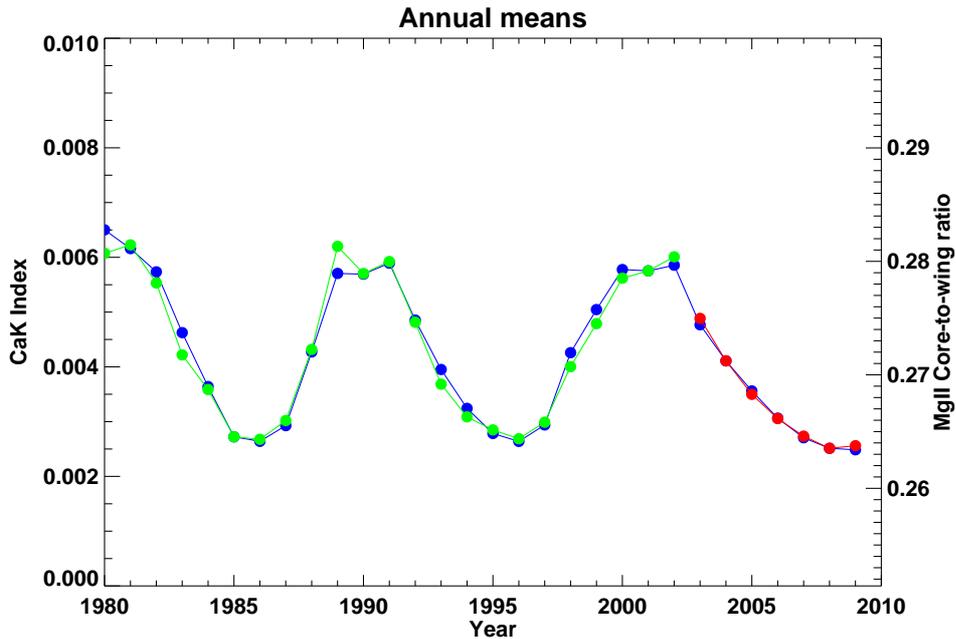}
\caption{Comparison between the CaK index (blue points) calculated in this work
and the Mg {\sc ii}~ core-to-wing ratio for the time interval 1980-2009. The Mg {\sc ii}~
index shown in this figure has been derived using the NOAA Mg {\sc ii}~ daily index (green
points) version 9.1 for the period
1980 to 2002, and the SORCE satellite data (red points) for the period 2003-present. Both these
time series are available from the National Geophysical Data Center website at
www.ngdc.noaa.gov/stp/SOLAR/ftpsolaruv.html. Although error bars are not shown in this plot
for reason of clarity, the separation between the CaK index and the corresponding
Mg {\sc ii}~ points is always less than the cumulative 1-$\sigma$ errors.
}
\label{mgii}
\end{center}
\end{figure}

Finally, visual inspection of Figure \ref{xcak} seems to suggest a subtle but systematic 
decreasing in the value of the CaK index, around minima of solar activity, from 1945 to present.
It is well established that chromospheric indices such as CaK index, Mg {\sc ii}~ core-to-wing ratio,
and Ly-$\alpha$ are highly correlated with the UV and EUV irradiance. The Mg {\sc ii}~ index, in 
particular, has been recently investigated to determine whether the spectral irradiance
below about 300 nm shows some secular changes. 
\inlinecite{2009A&A...501L..27F} has found evidence of a long-term trend in TSI, but not
in solar UV irradiance. The study of the long-term properties of the UV irradiance was
however limited to the investigation of the temporal behavior of the Mg {\sc ii}~ index over the last
three solar cycles, and confirmed by observations of Ly-$\alpha$ and \caii K
over the same time period. In this paper we have the opportunity to significantly extend this
time interval, by using our CaK index time series as a proxy for UV and EUV irradiance.
This is validated by the high correlation between our CaK index and the
Mg {\sc ii}~ core-to-wing ratio shown in Figure \ref{mgii}. Assuming that this correlation is 
preserved also for the previous cycles, shown in Figure \ref{xcak}, we must conclude that
the "quiet" magnetic Sun has slightly and systematically
reduced its UV irradiance over the past 70 years. 
A regression analysis using the CaK index values from Table 1 corresponding to years of minimum
in solar activity was performed to validate this conclusion. When only the last three or four
solar minimum values are included in this analysis, the significance 
(according to a Student's t-distribution) of the calculated
slope is around or well below 
the 95\% confidence level, for the four and three minima respectively. This result leads
to the conclusion that no significant changes have occurred in the quiet Sun UV irradiance
over the last 4 decades, as also shown by
other studies ({\it e.g.} \opencite{2009A&A...501L..27F} and references therein). However,
when the last six or seven solar minimum value are considered, a significant slope- at a
confidence level above 99.9\%- is found. The value of this slope is about 0.9/year,
indicating a variation of $\approx$20-25\% in the value of the "quiet" CaK index between solar
cycles 17/18 and 23/24. We are not able to calculate at present how much of this variation, 
if any, would produce changes in the quiet-Sun's output of UV. If such a variation can
be demonstrated, it could significantly impact the study of the possible effect
of the Sun on the Earth's climate, as discussed in \inlinecite{2006Natur.443..161F}.
We have repeated this analysis using the smoothed temporal behavior of the daily CaK
index instead of the annual means, and reached the same conclusions.
Further investigation and comparison with other data sets are required, however, 
to positively confirm this result.

\begin{acks}
This work has been supported by NSF grant to UCLA NSF ATM-0236682.
\end{acks}


\end{article} 
\end{document}